\documentclass{ws-ijmpb}

\begin{document}

\markboth{F. Benatti and R. Floreanini}
{Asymptotic Entanglement of Two Independent Systems in a Common Bath}

\catchline{}{}{}{}{}
%
%%%%%%%%%%%%%%%%%%%%%%%%%%%%%%%%%%%%%%%%%%%%%%%%%%%%%%%%%%%%%%%%%%%%

\title{ASYMPTOTIC ENTANGLEMENT OF TWO INDEPENDENT SYSTEMS IN A COMMON BATH}

\author{F. BENATTI}

\address{Dipartimento di Fisica Teorica, Universit\`a di Trieste,
\\
Strada Costiera 11, 34014 Trieste, Italy\\
and Istituto Nazionale di Fisica Nucleare, Sezione di Trieste\\
benatti@ts.infn.it}

\author{R. FLOREANINI}

\address{Istituto Nazionale di Fisica Nucleare, Sezione di Trieste,
\\
Dipartimento di Fisica Teorica, Universit\`a di Trieste,\\
Strada Costiera 11, 34014 Trieste, Italy\\
florean@ts.infn.it}

\maketitle

%\begin{history}
%\received{(Day Month Year)}
%\revised{(Day Month Year)}
%\accepted{(Day Month Year)}
%\comby{(xxxxxxxxxx)}
%\end{history}

\begin{abstract}
Two, non-interacting systems immersed in a common 
bath and evolving with a Markovian, completely positive dynamics
can become initially entangled via a purely noisy mechanism. 
Remarkably, for certain, phenomenologically relevant
environments, the quantum correlations can persist 
even in the asymptotic long-time regime.
\end{abstract}

%\keywords{Keyword1; keyword2; keyword3.}

\section{Introduction}

The simplest way to generate quantum correlations between two systems is 
via a suitable Hamiltonian coupling: optimization of entanglement production
can be achieved by a careful choice of the form 
of the interaction term.\cite{1}$^-$\cite{5}

When the two systems are immersed in an external bath, decoherence phenomena
usually occur, counteracting entanglement generation. 
These effects are clearly a curse in quantum information and in fact
various error correcting strategies have been devised in order to limit these
environment induced mixing enhancing phenomena.

However, an external environment can also provide a further, indirect coupling
between the two systems and therefore 
an additional mechanism to correlate them.\cite{6}$^-$\cite{9}
That the external environment can indeed generate entanglement has been
first established in exactly solvable models:\cite{6} there,
correlations between the two subsystems take place during a
short time transient phase, where the reduced dynamics
of the subsystems contains memory effects.

Remarkably, a similar phenomenon of entanglement production
may occur also in the
Markovian regime, through a purely noisy mechanism.\cite{10}$^-$\cite{12}$^,$\cite{28,29}
For the case of two, two-level systems immersed
in a common bath this phenomenon can be established by looking
at the eigenvalues of the partial transposed density matrix that
represents the two subsystem state.\cite{13,14} 
A sufficient condition for initial, 
environment induced entanglement generation can then be obtained:
it allows deriving a test on the entanglement power of the bath.\cite{10}

Nevertheless, this test is unable to determine the fate of the
initially created quantum correlations as time becomes large.
In order to discuss asymptotic entanglement, one has to
analyze directly the structure of the dissipative,
Markovian dynamics followed by the two systems, and
determine its equilibrium states.

In the following, we shall present such an investigation for 
a subsystem composed by a couple of identical two-level systems, evolving with
a completely positive dynamical semigroup. Being interested in discussing the
correlation power of the environment, we shall assume 
the two systems to be independent, without any mutual direct interaction.%
\footnote{In the same way, we shall also neglect any Lamb shift 
Hamiltonian contribution that might be generated by the presence of the
external bath.}
As we shall see, there exists a class of environments,
which is of relevance in phenomenological applications, that are
able not only to initially generate entanglement: they can continue to
enhance it even in the asymptotic long time regime.

\section{Master Equation}

We shall deal with a subsystem composed by two, identical,
non-interacting two-level systems, immersed in a common
heat bath. 
The derivation of a physically consistent master equation
for the reduced density matrix 
$\rho(t)\equiv{\rm Tr}_{\cal E}[\rho_{\rm tot}(t)]$, obtained by tracing
the total density matrix $\rho_{\rm tot}(t)$ over the bath degrees of freedom, 
is notoriously tricky, requiring an {\it a priori}
unambiguous separation between subsystem and environment.\cite{15}$^-$\cite{18}$^,$\cite{28}
Generally speaking, this distinction can be achieved when the correlations in the
environment decay much faster than the characteristic evolution time of the subsystem
alone. Then, in the limit of weak couplings, the changes in the
evolution of the subsystem occur on time scales that are very long, so large
that the details of the internal environment dynamics result irrelevant.

This situation is amenable to a precise mathematical treatment:\cite{19} as a result,
the two-system state $\rho(t)$
evolves in time according to a quantum dynamical
semigroup of completely positive maps, generated by a master equation
in Kossakowski-Lindblad form:\cite{19,20,21}
\begin{equation}
{\partial\rho(t)\over \partial t}= -i \big[{\cal H}_{\rm eff},\, \rho(t)\big]
 + {\cal L}[\rho(t)]\ .
\label{1}
\end{equation}
The unitary term depends on an effective Hamiltonian, containing
both the initial system Hamiltonian and suitable Lamb
contributions; in general, it can be decomposed as:
${\cal H}_{\rm eff}=H_{\rm eff}^{(1)}+H_{\rm eff}^{(2)}+H_{\rm eff}^{(12)}$.
The first two terms represent single system contributions,
\begin{equation}
H^{(1)}_{\rm eff}=\sum_{i=1}^3 H_i^{(1)} (\sigma_i\otimes {\bf 1})\ ,\quad
H^{(2)}_{\rm eff}=\sum_{i=1}^3 H_i^{(2)} ({\bf 1}\otimes \sigma_i)\ ,
\label{2}
\end{equation}
where $(\sigma_i\otimes{\bf 1})$, $({\bf 1}\otimes\sigma_i)$ 
are the basis operators pertaining to the two systems, respectively,
with $\sigma_i$, $i=1,2,3$ the Pauli matrices;
the third piece is a bath generated
direct two-system coupling term, that can be expressed as:
\begin{equation}
H_{\rm eff}^{(12)}=\sum_{i,j=1}^3 H^{(12)}_{ij}\, (\sigma_i\otimes\sigma_j)\ .
\label{3}
\end{equation}
The dissipative contribution ${\cal L}[\rho(t)]$ 
can be cast in Kossakowski form,\cite{20}
\begin{equation}
{\cal L}[\rho]=\sum_{\alpha,\beta=1}^6
C_{\alpha\beta} \bigg[ 
{\cal F}_\beta\, \rho\ {\cal F}_\alpha\, -\, {1\over 2}
\Bigl\{{\cal F}_\alpha {\cal F}_\beta\,,\,\rho\Bigr\}\bigg]\ ,
\label{4}
\end{equation}
using the hermitian, traceless, matrices ${\cal F}_\alpha$
that coincide with the first system basis operators 
$(\sigma_\alpha\otimes{\bf 1})$ for $\alpha=1,2,3$, while
reproducing the second system basis operators
$({\bf 1}\otimes \sigma_{\alpha-3})$ for $\alpha=4,5,6$.
The Kossakowski matrix $C_{\alpha\beta}$ is a $6\times 6$ matrix,
which is non-negative, thus guaranteeing the complete positivity
of the reduced dynamics. Using the splitting of the indices
introduced above, it can be conveniently written as
\begin{equation}
C=\pmatrix{
  {\cal A} & {\cal B} \cr
  {\cal B}^{\dagger} & {\cal C}}\ ,
\label{5}
\end{equation}
in terms of $3\times 3$ matrices ${\cal A}={\cal A}^\dagger$, 
${\cal C}={\cal C}^\dagger$ and ${\cal B}$.
This decomposition carries a direct
physical interpretation. Indeed, the pieces in (\ref{4})
containing the diagonal contributions 
${\cal A}$ and ${\cal C}$ correspond
to noise terms that affect the first, respectively the second, 
system in absence of the other. 
On the contrary, the pieces depending on $\cal B$ encode
environment generated dissipative couplings between 
the two, otherwise independent, systems.

In order to obtain a more explicit expression for the dynamics, 
it is convenient to decompose the $4\times4$ density matrix $\rho(t)$
along the Pauli matrices:
\begin{equation}
\rho(t)={1\over4}\bigg[{\bf 1}\otimes{\bf 1}+
\sum_{i=1}^3 \rho_{0i}(t)\ {\bf 1}\otimes\sigma_i
+ \sum_{i=1}^3 \rho_{i0}(t)\ \sigma_i\otimes{\bf 1}
+ \sum_{i,j=1}^3 \rho_{ij}(t)\ \sigma_i\otimes\sigma_j\bigg]\ ,
\label{6}
\end{equation}
where the coefficients $\rho_{0i}(t)$, $\rho_{i0}(t)$, $\rho_{ij}(t)$ are all real.
Substitution of this expansion in the master equation (\ref{1}) allows deriving
the corresponding evolution equations for the above components of $\rho(t)$.
As mentioned in the Introduction, we are interested in studying possible entanglement
production through the purely dissipative action of the environment;
in the following, we shall therefore ignore the Hamiltonian pieces
and concentrate on the study of the effects induced by the
dissipative part ${\cal L}[\,\cdot\,]$ in (\ref{4}).%
\footnote{In other terms, we limit our analysis to baths for which
the induced two-system Hamiltonian coupling in (\ref{3}) is vanishingly
small or alternatively does not give rise to temperature dependent 
entanglement phenomena. This situation is rather common in phenomenological
applications; for concrete examples, see Refs.[11,12].}

Although the general case can be similarly treated,
for sake of simplicity we shall here limit our considerations to baths
for which the submatrices in (\ref{5}) are all equal:
${\cal A}={\cal B}={\cal C}$. This choice of the
Kossakowski matrix, although special, is nevertheless of great 
phenomenological relevance since it is adopted in the analysis 
of the phenomenon of resonance fluorescence.\cite{22,23} 
In addition, it is precisely a dissipative term of this type that
describes the interaction of two
atoms with a set of weakly coupled external quantum fields, 
in the limit of a vanishing spatial atom separation.\cite{11,12}

In such a situation, the form of the dissipative contribution in
(\ref{4}) simplifies so that the evolution equation
can be rewritten as
\begin{equation}
{\partial\rho(t)\over \partial t}=
\sum_{i,j=1}^3
{\cal A}_{ij}\bigg[
\Sigma_j\,\rho(t)\,\Sigma_i
-\frac{1}{2}\Big\{\Sigma_i\Sigma_j\,,\,\rho(t)\Big\}
\bigg]\ ,
\label{7}
\end{equation}
in terms of the following symmetrized two-system operators
\begin{equation}
\Sigma_i=\sigma_i\otimes{\bf 1} + {\bf 1}\otimes\sigma_i\ ,
\quad i=1,2,3\ .
\label{8}
\end{equation}
One easily checks that these operators 
obey the same $su(2)$ Lie algebra of the Pauli matrices;
further, together with
\begin{equation}
S_{ij}=\sigma_i\otimes\sigma_j + \sigma_j\otimes\sigma_i\ ,
\quad i,j=1,2,3\ ,
\label{9}
\end{equation}
they form a closed algebra under matrix multiplication, 
whose explicit expression is collected
in the Appendix.

It is now convenient to decompose the hermitian matrix ${\cal A}_{ij}$
into its real and imaginary parts,
\begin{equation}
{\cal A}_{ij}=A_{ij}+i\sum_{k=1}^3\varepsilon_{ijk} B_k\ ,
\label{10}
\end{equation}
with $A_{ij}$ real symmetric and $B_i$ real. Inserting this 
in (\ref{7}) and using the decomposition (\ref{6}),
a straightforward but lengthy calculation allows derive 
the following evolution equations for the components of $\rho(t)$:
\begin{eqnarray}
\nonumber
&&{\partial\rho_{0i}(t)\over \partial t}=-2A\rho_{0i}(t)
+2\sum_{k=1}^3 \Big[A_{ik}\rho_{0k}(t) -\rho_{ik}(t) B_k\Big]
+2(2+\tau) B_i\ ,\\
\nonumber
&&{\partial\rho_{i0}(t)\over \partial t}=-2A\rho_{i0}(t)
+2\sum_{k=1}^3 \Big[A_{ik}\rho_{k0}(t) -\rho_{ki}(t) B_k\Big]
+2(2+\tau) B_i\ ,\\
\nonumber
&&{\partial\rho_{ij}(t)\over \partial t}=
-4A\big[\rho_{ij}(t)+\rho_{ji}(t)\big]
+2\sum_{k=1}^3\Big[A_{ik}\rho_{kj}(t)+A_{jk}\rho_{ik}(t)\Big]-4A_{ij}\tau\\
\nonumber
&&\hskip 1.6cm +4\sum_{k=1}^3\Big[A_{ik}\rho_{jk}(t)+A_{jk}\rho_{ki}(t)\Big]
+4\Big[A\tau - \sum_{k,l=1}^3 A_{kl}\rho_{lk}(t)\Big]\,\delta_{ij}\\
\nonumber
&&\hskip 1.6cm +2\Big[B_i\rho_{j0}(t) + B_j\rho_{0i}(t)\Big]
+4\Big[B_i\rho_{0j}(t) + B_j\rho_{i0}(t)\Big]\\
%\nonumber
&&\hskip 1.6cm -2\sum_{k=1}^3 B_k \big[\rho_{0k}(t) + \rho_{k0}(t)\big]\,\delta_{ij}\ .
\label{11}
\end{eqnarray}
In these formulas, the parameter $A$ represents
the trace of $A_{ij}$ 
while the quantity $\tau\equiv\sum_{i=1}^3\rho_{ii}$ 
that of the submatrix $\rho_{ij}$. By taking the trace
of both sides of the last equation above,
one discovers that $\tau$
is a constant of motion.%
\footnote{By analyzing the structure of a general evolution
equation with the dissipative term as in (\ref{4}), one can show that 
this is the case only when the condition
${\cal A}={\cal B}={\cal C}$ is satisfied. 
This result is also related
to the existence of multiple equilibrium states; see below.}
Nevertheless,
the value of $\tau$ can not be chosen arbitrarily; the requirement
of positivity of the initial density matrix $\rho(0)$ readily
implies: \hbox{$-3\leq\tau\leq 1$}.

\section{Environment Induced Entanglement Generation}

Using the explicit form (\ref{11}) for the master equation
derived in the previous Section, one can now
investigate whether an external environment
can actually entangle the two independent systems.
Since we are dealing with a couple of two-level systems,
this can be achieved with the help of the 
partial transposition criterion:\cite{13,14}
a state $\rho(t)$ results entangled at time $t$
if and only if the operation of partial transposition
does not preserve its positivity.

We shall first discuss the possibility of entanglement creation
at the beginning of the evolution: if a bath is not
able to initially entangle the two systems, it will hardly
do so in the limit of large times.
A simple strategy to ascertain entanglement creation is
as follows: assume the initial state to be pure and separable,
{\it i.e.} $\rho(0)=\vert \varphi\rangle\langle \varphi\vert\otimes 
\vert \psi\rangle \langle \psi\vert$, and then find out
whether in the neighborhood of $t=\,0$ the dynamics in (\ref{11}) 
is able to make negative an initially zero eigenvalue 
of the partially transposed density matrix $\tilde\rho(t)$
(note that $\tilde\rho(0)\equiv\rho(0)$ for the chosen initial state). 
This amounts to study the behavior of the time derivative
$\partial_t\tilde\rho(0)$, that can be explicitly obtained by
taking the partial transposition of both sides of (\ref{7})
(or equivalently of the system of equations in (\ref{11})).

In this way, one finds that the dynamics generated
by the equations in (\ref{7}) can indeed
entangle the two subsystems,
provided at least one of the coefficients $B_i$ in the
Kossakowski matrix ${\cal A}_{ij}$ in (\ref{10}) is nonvanishing
(see Ref.[10] for further details). 
This is a generic property of the Markovian dynamics
in (\ref{7}): entanglement is generated as soon as
$t>0$.

In order to study the fate of this initially produced
correlations as time becomes large, one needs to
analyze the ergodic properties of the semigroup evolution
generated by (\ref{7}).
On general grounds, one expects that the effects 
of decoherence and dissipation that
counteract entanglement production
be dominant at large times, so that 
no entanglement is left at the end.
However, as we shall explicitly see, there are
cases for which the environment induced entanglement 
creation never stops as time flows, allowing at the end
the presence of entangled equilibrium states.

The system of first order differential equations in (\ref{11}) naturally splits
into two independent sets, involving the symmetric,
$\rho_{(0i)}=\rho_{0i}+\rho_{i0}$, $\rho_{(ij)}=\rho_{ij}+\rho_{ji}$,
and antisymmetric,
$\rho_{[0i]}=\rho_{0i}-\rho_{i0}$, $\rho_{[ij]}=\rho_{ij}-\rho_{ji}$,
variables. By examining the structure 
of the two sets of differential
equations, one can conclude that the antisymmetric variables
involve exponentially decaying factors, so that they vanish for
large times. Then, recalling the definitions in (\ref{8}) and (\ref{9}),
the study of the equilibrium states $\hat\rho$ of the
evolution (\ref{7}) can be limited to density matrices of the form:
\begin{equation}
\hat\rho={1\over4}\bigg[{\bf 1}\otimes{\bf 1}+
\sum_{i=1}^3 \hat\rho_{i}\, \Sigma_i
+ \sum_{i,j=1}^3 \hat\rho_{ij}\,  S_{ij}\bigg]\ ,
\label{12}
\end{equation}
with $\hat\rho_{ij}=\hat\rho_{ji}$.

These states obey the equilibrium condition $\partial_t\hat\rho=\,0$,
and therefore annihilates the r.h.s. of all equations in (\ref{11}).
By direct inspection, one finds that these conditions are invariant
under linear orthogonal transformations that
act on both the coefficients $B_i$, $A_{ij}$
and the components $\hat\rho_i$, $\hat\rho_{ij}$ 
of the density matrix $\hat\rho$. Then, without loss of generality,
for the purpose of identifying the asymptotic states,
one can take the real part of ${\cal A}_{ij}$ to be diagonal, {\it i.e.}
$A_{ij}=\lambda_i\,\delta_{ij}$; the general case can always be recovered
at the end by undoing the orthogonal transformation that has brought
$A_{ij}$ in diagonal form. 

Further, in order to simplify the exposition,
we shall assume the vector of components $B_i$ to be directed along
the third axis, so that only the component $B_3\equiv B$ will be
nonvanishing. Then, positivity of ${\cal A}_{ij}$ readily implies:
$\lambda_i\geq 0$, $i=1,2,3$ and $B^2\leq \lambda_1 \lambda_2$.

The approach to equilibrium of semigroups whose generator
is of the generic Kossakowski-Lindblad form has been studied
in general and some rigorous mathematical results are
available.\cite{24,16} We shall present such results
by adapting them to the case of the evolution
generated by the equation (\ref{7}).

First of all, one notice that in the case
of a finite dimensional Hilbert space,
there always exists at least one stationary state $\hat\rho_0$.%
\footnote{This can be understood by recalling that in
finite dimensions the ergodic average of the action of 
a completely positive one-parameter 
semigroup on any initial state always exists:
the result is clearly a stationary state.}
Let us now introduce the operators 
$V_i=\sum_{j=1}^3{\cal A}^{1/2}_{ij}\, \Sigma_j$
(recall that ${\cal A}$ is non-negative), so that 
the r.h.s. of (\ref{7})
can be rewritten in the so-called diagonal form:
\begin{equation}
{\cal L}[\rho]\!=\!\!
\sum_{i,j=1}^3
\bigg[
V_j\,\rho\,V_i^\dagger
-\frac{1}{2}\Big\{V_i^\dagger V_j\,,\,\rho\Big\}
\bigg]\ .
\label{13}
\end{equation}
When the set ${\cal M}$ formed by all operators that commute with the 
linear span of $\{V_i,\ V_i^\dagger,\ i=1,2,3\}$
contains only the identity, one can show that 
the stationary state $\hat\rho_0$ 
is unique, and of maximal rank. 
On the other hand, when there are several stationary states,
they can be generated in a canonical way from a $\hat\rho_0$ with
maximal rank using the elements of the set $\cal M$.

In the present case, $\cal M$ contains the operator
$S\equiv\sum_{i=1}^3 S_{ii}$, besides the identity;
indeed, with the help of the algebraic
relations collected in the Appendix, one immediately
finds: $[S,\ \Sigma_i]=\,0$.
Out of these two elements of $\cal M$, one can now construct
two mutually orthogonal projection operators:%
\footnote{One easily checks that $P$ is the projection operator
on one of the maximally entangled Bell states.}
\begin{equation}
P={1\over4}\bigg[{\bf 1}\otimes{\bf 1}-\frac{S}{2}\bigg]\ ,\qquad
Q=1-P\ .
\label{14}
\end{equation}
Then, one can show that any given initial state $\rho(0)$ 
will be mapped by the evolution (\ref{7}) into the following
equilibrium state:
\begin{equation}
\rho(0)\to\hat\rho=\frac{P\, \hat\rho_0\, P}{{\rm Tr}\big[P\, \hat\rho_0\, P\big]}\
{\rm Tr}\big[P\, \rho(0)\big]
+\frac{Q\, \hat\rho_0\, Q}{{\rm Tr}\big[Q\, \hat\rho_0\, Q\big]}\
{\rm Tr}\big[Q\, \rho(0)\big]\ .
\label{15}
\end{equation}
That this state is indeed stationary can be easily proven 
by recalling that $P$ and $Q$ commute with $\Sigma_i$, $i=1,2,3$,
and thus with the $V_i$ as well;
therefore, ${\cal L}[\hat\rho]=\,0$, for any $\rho(0)$,
as a consequence of ${\cal L}[\hat\rho_0]=\,0$.

The problem of finding all invariant states of the dynamics (\ref{7})
is then reduced to that of identifying a stationary state
$\hat\rho_0$ with all eigenvalues nonzero.
Although in principle this amounts in solving a linear algebraic equation,
in practice it can be rather difficult for
general master equations of the form (\ref{1}).
Nevertheless, in the case at hand, the problem can be explicitly solved,
yielding:
\begin{equation}
\hat\rho_0={1\over4}\bigg[{\bf 1}\otimes{\bf 1}+
M\, \Sigma_3
- N\, \big(S_{11}-S_{22}\big) +R\, S_{33}\bigg]\ ,
\label{16}
\end{equation}
with
\begin{eqnarray}
&&M=\frac{2\, B}{\lambda_1+\lambda_2}\ ,\\
&&N=\frac{(\lambda_1-\lambda_2)\, B^2}{2(\lambda_1+\lambda_2)
(\lambda_1\lambda_2 + \lambda_1\lambda_3 +\lambda_2\lambda_3)}\ ,\\
&&R=\frac{(\lambda_1+\lambda_2+4\,\lambda_3)\, B^2}{2(\lambda_1+\lambda_2)
(\lambda_1\lambda_2 + \lambda_1\lambda_3 +\lambda_2\lambda_3)}\ .
\label{19}
\end{eqnarray}
Note that the previously mentioned positivity conditions on the parameters
$\lambda_i$, $i=1,2,3$ and $B$, put restrictions on the components
of $\hat\rho_0$ given above; in particular, one has: $0\leq2\,R\leq1$,
$M^2\leq 2R$, $M^2+4N^2\leq1$, and the upper limits can be reached
only when $\lambda_1=\lambda_2$.%
\footnote{These inequalities also guarantee that $\hat\rho_0$ as
given in (\ref{16}) is indeed a state, {\it i.e.} that all its
eigenvalues are non-negative.}

Inserting this result in the expression (\ref{15}) allows deriving
the expression of the set of all equilibrium states of the
dynamics (\ref{7}); as expected, they take the symmetric form
of (\ref{12}), with the nonvanishing components
given by:
\begin{eqnarray}
&&\hat\rho_3=\frac{3+\tau}{3+2R}\,M\ ,\\
&&\hat\rho_{11}=\frac{(1+2N)\tau+2(3N-R)}{2(3+2R)}\ ,\\
&&\hat\rho_{22}=\frac{(1-2N)\tau-2(3N+R)}{2(3+2R)}\ ,\\
&&\hat\rho_{33}=\frac{4R+(1+2R)\tau}{2(3+2R)}\ .
\label{23}
\end{eqnarray}
These stationary density matrices depend on the initial condition $\rho(0)$
only through the value of the parameter $\tau$, that as already mentioned
is a constant of motion for the dynamics in (\ref{7}).

Now that we have completely classified the stationary states, one can
study their properties, in particular with respect to quantum correlations.
It turns out that $\hat\rho$ is in general entangled.

To explicitly show this, one can as before act with the operation of
partial transposition on $\hat\rho$ to see whether negative eigenvalues
are present. Alternatively, one can resort to one of the available
entanglement measures and concurrence appears here to be 
the more appropriate: its value ${\cal C}[\rho]$ ranges from zero, for separable
states, to one, for fully entangled states.\cite{25}$^-$\cite{27}
In the case of the state $\hat\rho$ above, 
one finds
\begin{equation}
{\cal C}[\hat\rho]={\rm max}\Bigg\{{\big(2+\Delta\big)\over2\big(3+2R\big)}\,
\bigg[ {4R-3\Delta\over 2+\Delta} -\tau\bigg],\ 0\Bigg\}\ ,
\label{24}
\end{equation}
where
\begin{equation}
\Delta=\Big[\big(1-2R\big)^2+4(2R-M^2)\Big]^{1/2}\ .
\label{25}
\end{equation}
The expression in (\ref{24}) is indeed nonvanishing, provided we start with
an initial state $\rho(0)$ for which
\begin{equation}
\tau< {4R-3\Delta\over 2+\Delta}\ .
\label{26}
\end{equation}
The concurrence depends linearly on the initial parameter $\tau$;
it assumes its maximum value ${\cal C}[\hat\rho]=1$ when $\tau=-3$,
as for the state $P$ in (\ref{14}), 
and reaches zero at $\tau=(4R-3\Delta)/(2+\Delta)\leq 1$.

This result is remarkable, since it implies that the dynamics
in (\ref{7}) not only can initially generate entanglement: 
it can continue to enhance it even in the asymptotic long time
regime. In other terms, prepare the two atoms in a separable 
state at $t=\,0$; then, provided the condition (\ref{26}) is satisfied,
their long time equilibrium state will turn out to be entangled.

Entanglement enhancement is not
limited though to initially separable states: one can 
easily check that the phenomenon 
of entanglement production through a purely noise mechanism
takes place also when
the initial state $\rho(0)$ already has a non-vanishing concurrence.
As an example, let us consider the following 
initial state, built out of
the two projector operators introduced in (\ref{14}):
\begin{equation}
\rho(0)={s\over3} Q+(1-s)P \ ;
\label{27}
\end{equation}
it interpolates between the completely mixed (separable) state
obtained for $s=3/4$ and the totally entangled state $P$.
Provided $s<1/2$, this state is entangled, with
${\cal C}[\rho(0)]=1-2s$. The difference
in concurrence as 
this initial state evolves to its corresponding asymptotic
one $\hat\rho$ turns out to be
\begin{equation}
{\cal C}[\hat\rho]-{\cal C}[\rho(0)]=2s\bigg[1-{2+\Delta\over 3+2R}\bigg]\ ,
\label{28}
\end{equation}
which is indeed non vanishing. As a final remark, notice that this
enhancement in concurrence vanishes as $s$ approaches zero; 
in other terms, the maximally entangled state $P$ can never
be reached as an asymptotic state unless one already
starts with it at $t=\,0$: $P$ results an isolated fixed point
of the dynamics generated by (\ref{7}).

\section*{Appendix}

We collect here the algebraic relations obeyed by the nine hermitian, traceless
matrices $\Sigma_i=\sigma_i\otimes{\bf 1} + {\bf 1}\otimes\sigma_i$
and $S_{ij}=\sigma_i\otimes\sigma_j + \sigma_j\otimes\sigma_i$,
$i,j=1,2,3$, introduced in (\ref{8}) and (\ref{9}). As mentioned in
the text, altogether they form a closed algebra under matrix multiplication.
In fact, using $\sigma_i\sigma_j=\delta_{ij}+i\sum_{k=1}^3\varepsilon_{ijk}\, \sigma_k$,
a direct computation yields:
\begin{eqnarray}
\nonumber
&&\Sigma_i\, \Sigma_j=2\,\delta_{ij}\, {\bf 1}\otimes{\bf 1}
+i\sum_{k=1}^3\varepsilon_{ijk}\,\Sigma_k+S_{ij}\ ,\\
\nonumber
&&S_{ij}\, \Sigma_k=\delta_{ik}\, \Sigma_j + \delta_{jk}\, \Sigma_i
+i\sum_{l=1}^3\varepsilon_{ikl}\, S_{lj} + i\sum_{l=1}^3\varepsilon_{jkl}\, S_{il}\ ,\\
\nonumber
&&\Sigma_k\, S_{ij}=\delta_{ik}\, \Sigma_j + \delta_{jk}\, \Sigma_i
-i\sum_{l=1}^3\varepsilon_{ikl}\, S_{lj} - i\sum_{l=1}^3\varepsilon_{jkl}\, S_{il}\ ,\\
\nonumber
&&S_{ij}\, S_{kl}= 2\big(\delta_{ik}\,\delta_{jl}+\delta_{il}\,\delta_{jk}\big)
{\bf 1}\otimes{\bf 1}
+i\sum_{r=1}^3\Big(\delta_{ik}\varepsilon_{jlr}+\delta_{jk}\varepsilon_{ilr}
+\delta_{il}\varepsilon_{jkr}+\delta_{jl}\varepsilon_{ikr}\Big)\, \Sigma_r\\
\nonumber
&&\hskip 1.5cm -\Big(2\,\delta_{ij}\,\delta_{kl}-\delta_{ik}\,\delta_{jl}
-\delta_{il}\,\delta_{jk}\Big)\, S
+2\Big(\delta_{ij}\, S_{kl} + \delta_{kl}\, S_{ij}\Big)\\
\nonumber
&&\hskip 1.5cm -\delta_{ik}\, S_{jl} -\delta_{il}\, S_{jk}-\delta_{jk}\, S_{il}
-\delta_{jl}\, S_{ik}\ ,
\label{a1}
\end{eqnarray}
where $S=\sum_{r=i}^3 S_{ii}$. From these relations, one immediately sees
that the commutant of the linear span of the set $\{\Sigma_i\!:\ i,j=1,2,3 \}$ 
contains two elements, ${\bf 1}\otimes{\bf 1}$
and $S$. As explained in \hbox{Section 3}, this result allows classifying all stationary
states of the open dynamics generated by the evolution equations in (\ref{7}).

\end{document}